\documentclass[letterpaper, 10 pt, journal, twoside]{IEEEtran}

\IEEEoverridecommandlockouts                              

\usepackage{ifpdf}
\usepackage{amsmath}
\usepackage{amsfonts}
\usepackage{mathrsfs}
\usepackage{color}
\usepackage{graphicx}
\usepackage{subfigure}
\usepackage{epstopdf}
\usepackage{lineno}
\usepackage{cite}
\usepackage{enumerate}
\usepackage{algorithm, algorithmic}

\title{\LARGE \bf
Distributed Localization in Wireless Sensor Networks Under Denial-of-Service Attacks}

\newtheorem{theorem}{\textbf{Theorem}}

\newtheorem{definition}{\textbf{Definition}}
\newtheorem{example}{\textbf{Example}}

\newtheorem{remark}{Remark}

\author{Lei Shi, Qingchen Liu, Jinliang Shao, and Yuhua Cheng,~\IEEEmembership{Senior Member, IEEE}
\thanks{This research was supported in part by the European Union's Horizon 2020 research and innovation programme under the Marie Sk{\l}odowska-Curie grant agreement (754462), the National Science Foundation of China (U1830207, 61772003, 61903066), the China Postdoctoral Science Foundation (2017M612944, 2018T110962), and the National Key R\&D Program of China (2017YFC1501005, 2018YFC1505203).}
\thanks{L. Shi, Jinliang Shao and Y. Cheng are with the School of Automation Engineering, University of Electronic Science and Technology of China, Sichuan, 611731, China (email: shilei@std.uestc.edu.cn, jinliangshao@uestc.edu.cn, chengyuhua\_auto@uestc.edu.cn)}%
\thanks{Qingchen Liu is with the Chair of Information-Oriented Control, Technical University of Munich, Munich, 80869, Germany (email: qingchen.liu@tum.de)}%
}

\graphicspath{{figures/}}

\begin{document}

\maketitle
\thispagestyle{empty}
\pagestyle{empty}

\begin{abstract}
In this paper, we study the problem of localizing the sensors' positions in presence of denial-of-service (DoS) attacks. We consider a general attack model, in which the attacker action is only constrained through the frequency and duration of DoS attacks. We propose a distributed iterative localization algorithm with an abandonment strategy based on the barycentric coordinate of a sensor with respect to its neighbors, which is computed through relative distance measurements. In particular, if a sensor's communication links for receiving its neighbors' information lose packets due to DoS attacks, then the sensor abandons the location estimation. When the attacker launches DoS attacks, the AS-DILOC algorithm is proved theoretically to be able to accurately locate the sensors regardless of the attack strategy at each time. The effectiveness of the proposed algorithm is demonstrated through simulation examples.
\end{abstract}


\section{INTRODUCTION}

\IEEEPARstart{L}{ocalization}, which is a fundamental problem in wireless sensor networks, has attracted extensive attention in the past few decades. In some networks with centralized information structures, such as agricultural monitoring networks and road traffic monitoring networks, it is very convenient to use centralized localization method because the measurement data of all nodes in the network are collected in a central processor \cite{Shang2004Loc,Pal2010Loc}. However, as described in \cite{Mao2007Wire}, the centralized method is usually not suitable for large-scale sensor networks, and it requires higher computational complexity and lower reliability than the distributed method due to the accumulated information loss caused by the multi hop transmissions on the wireless network.

In 2009, Khan \emph{et al.} \cite{Khan2009Distributed} proposed a distributed iterative localization (DILOC) algorithm based on barycentric coordinates representation, where there are anchors, i.e., the nodes with known accurate locations, and sensors to be localized. The remarkable feature of the DILOC algorithm is that the sensor localization is represented by an iterative process with matrix-vector form through relative distance measurements. It has been proved that the DILOC algorithm can converge to the exact locations of sensors. Based on the framework of DILOC, the localization problem in different scenarios have been studied. In \cite{Khan2009Diland}, the DILOC algorithm was utilized to achieve precise localization in a noisy environment. Deghat \emph{et al.} \cite{Deghat2011Distributed} established an algebraic condition for the DILOC algorithm to achieve global convergence in a finite-time horizon. The global convergence with measurement noise and local convergence with least squares solutions of the DILOC algorithm were analyzed in \cite{Zhu2011Distributed}. Based on three types of hybrid measurements, including distance, azimuth and relative position, Lin \emph{et al.} \cite{Lin2017Distributed} presented a necessary and sufficient graphic condition for sensor localization. In addition, the reference \cite{Huang2018Asynchronous} proposed a more general iterative form for the DILOC algorithm by adding a gain parameter based on consensus of multi-agent systems (e.g., see \cite{Jadbabaie2003Coodi,Olfati2004Consensus}). This consensus-based iterative form includes the iterative form in \cite{Khan2009Distributed} as a special case.

Network security issues have gradually become the focus of public attention \cite{Mo2011C,Pasqualetti2013Attack,Fawzi2014Secure}. In recent years, various cyber attacks have appeared with different purposes. A common type of network attack is called denial of service (DoS), whose purpose is to block communication, thereby forcing network resources unavailable to its target users, i.e., packet loss. In most previous studies (e.g., \cite{DeBruhl2011Digital,Xiong2007Stabili}), packet loss was always assumed to follow a probability distribution. In fact, it is not feasible for a defender to assume that packet loss caused by attackers follows a given probability distribution \cite{Persis2015Input}. With this in mind, the reference \cite{Persis2015Input} proposed a general attack model without any assumptions on the underlying attack strategy, where the attacker action is only constrained through the frequency and duration of DoS attacks.

To the best of our knowledge, most work based on the DILOC algorithm is carried out in a perfect communication environment \cite{Khan2009Distributed,Khan2009Diland,Deghat2011Distributed,Zhu2011Distributed,Lin2017Distributed,Huang2018Asynchronous}. It is of practical significance to analyze the localization problem of wireless sensor networks under DoS attacks, which is also the research motivation of this paper. According to the general DoS attacks model proposed in \cite{Persis2015Input}, we first uses a simulation example to show that the classical DILOC algorithm may not achieve localization, while the attackers can attack the communication network in a changing active time period. Then, we propose a new consensus-based distributed iterative localization algorithm using an abandonment strategy, which is called the AS-DILOC algorithm. By introducing the composition of binary relations and casting the iteration of the AS-DILOC algorithm as a convergence problem of the product of infinite sub-stochastic matrices, we prove that the AS-DILOC algorithm converges globally to the accurate locations of sensors in presence of DoS attacks.

The outline of this paper is shown as follows. Section II introduces some preliminary knowledge. Section III presents
the AS-DILOC algorithm under DoS attacks in detail. In Section IV, the global
convergence of the AS-DILOC algorithm is theoretically
proved. The paper is concluded in Section V.

\section{Preliminaries}\label{section:2}

\subsection{Notations}\label{section:2.1}

The set of natural numbers is denoted by $\mathbb{N}$. $I_{n}$ is an $n$-order identity matrix. Let \textbf{0} be compatible dimensions of zeros, and $\mathbf{1}$ be an all $1$'s column vector. $M(i,:)$ denotes the $i$th row of matrix $M$. $\det[M]$ represents the determinant of matrix $M$. For a nonnegative matrix $M\in\mathbb{R}^{n\times n}$, its infinite norm is represented as $\|M\|_{\infty}=\max\big\{\sum_{j=1}^{n}M_{ij}: i=1,2,\ldots,n\big\}$. A nonnegative matrix $M$ is row-stochastic if $M\mathbf{1}=\mathbf{1}$, and it is sub-stochastic if $M\mathbf{1}\leq\mathbf{1}$. Let $\prod_{w=1}^sM_w=M_sM_{s-1}\cdots M_1$ represent the left products of matrices $M_i, i=1,2,\ldots,s$.

\begin{definition}[\cite{Rockafellar1970Convex}]\label{definition:1}
The convex hull of the set $\mathscr{X}=\{x_{1},x_2,\ldots,x_{n}\}$ is the minimum convex set containing all points $x_{i},i=1,2,\ldots,n$, represented by $Co\{\mathscr{X}\}=\{\sum^{n}_{i=1}a_{i}x_{i}: a_{i}\geq0, \sum^{n}_{i=1}a_{i}=1\}$.
\end{definition}

\subsection{Graph theory}\label{section:2.3}

Let $\mathscr{G}=\{\mathscr{V},\mathscr{E}\}$ represent a digraph containing a node set $\mathscr{V}=\{1,2,\ldots,n\}$ and an arc set $\mathscr{E}$. An arc with initial node $i$ and terminal node $j$ is represented by $(i,j)$. There are no self-loops, namely, such arcs as $(v_i,v_i)$, $i=1,2,\ldots,n$. A directed path from nodes $i_1$ to $i_z$ in $\mathscr{G}$ is a finite non-null edge sequence $\mathscr{P}_{i_1\rightarrow i_z}=(i_1,i_2)(i_2,i_3)\cdots (i_{z-1},i_z)$, in which $i_1,i_2,\ldots,i_z\in\mathscr{V}$ are different from each other. The directed distance from $i_1$ to $i_z$ is indicated as $d_{i_1\rightarrow i_z}$, which is the number of arcs in the shortest path from $i_1$ to $i_z$.

\begin{definition}[\cite{Rosen2012Discrete}]\label{difinition:2}
Let $\mathscr{E}_{1}$ and $\mathscr{E}_{2}$ be relations on a set $\mathscr{V}$. The composition of $\mathscr{E}_{1}$ and $\mathscr{E}_{2}$ is the relation consisting of order pairs $(b,d)$, where $b,d\in\mathscr{V}$, and for which there exists $c\in\mathscr{V}$ such that $(b,c)\in\mathscr{E}_{1}$ and $(c,d)\in\mathscr{E}_{2}$. The composition of $\mathscr{E}_{1}$ and $\mathscr{E}_{2}$ is represented by $\mathscr{E}_{1}\circ\mathscr{E}_{2}$.
\end{definition}

\subsection{Barycentric coordinate representation}\label{section:2.3}

\begin{figure}[htb]
  \centering
    \includegraphics[width=1.2in]{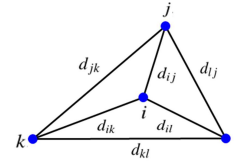}
  \caption{An example of node $i$ lying in the convex hull of nodes $j,k,l$.}\label{fig1}
\end{figure}

The barycentric coordinate describes the relative position of a node relative to other nodes. For four nodes $i,j,k,l$ in the 2-dimension Euclidean space, the coordinates are expressed by $p_i,p_j,p_k,p_l$, respectively. The barycentric coordinates of point $i$ relative to $j,k,l$ are $a_{ij},a_{ik},a_{il}$, satisfying
\begin{equation}\label{sys:1}
p_i=a_{ij}p_j+a_{ik}p_k+a_{il}p_l,
\end{equation}
where $a_{ij}+a_{ik}+a_{il}=1$. In particular, the barycentric coordinates can be expressed as the ratio of regions between specified triangles. For instance, in Fig. \ref{fig1}, the barycentric coordinates are given by
\begin{equation*}
\begin{aligned}
a_{ij}=\frac{S_{\triangle ikl}}{S_{\triangle jkl}}, \ a_{ik}=\frac{S_{\triangle ilj}}{S_{\triangle jkl}}, \ a_{il}=\frac{S_{\triangle ljk}}{S_{\triangle jkl}},
\end{aligned}
\end{equation*}
where $S_{\triangle ikl},S_{\triangle ilj},S_{\triangle ljk},S_{\triangle jkl}$ are the areas of the corresponding triangles $\triangle ikl,\triangle ilj,\triangle ljk,\triangle jkl$, which can be calculated with Euclidean distance measurements between nodes by \emph{Cayley-Menger} determinant \cite{Sippl1986Cayley}, i.e.,
\begin{equation*}
S^2_{\triangle ikl}=-\frac{1}{16}\det\begin{bmatrix}
                                                        0 & 1 & 1 & 1 \\
                                                        1 & 0 & d^2_{ik} & d^2_{il} \\
                                                        1 & d^2_{ki} & 0 & d^2_{kl} \\
                                                        1 & d^2_{li} & d^2_{lk} & 0 \\
\end{bmatrix},
\end{equation*}
in which $d_{ik},d_{il},d_{lk}$ are Euclidean distance measurements among nodes $i,l,k$.

\section{Problem formulation}\label{section:3}

\subsection{DILOC algorithm}\label{section:3.1}

Consider a wireless sensor network with a node set $\Theta=\{1,2,\ldots,n\}$ in the 2-dimension Euclidean space. A node is called an anchor if its location is known, and a node is called a sensor if its location is unknown and to be determined. As shown in \cite{Khan2009Distributed}, at least three anchors that are not in a straight line are necessary to locate any sensor with unknown location in the 2-dimension Euclidean space. Therefore, without loss of generality, we let $\Xi\!=\!\{1,2,3\}$ and $\Omega\!=\!\{4,5,\ldots,n\}$ represent the set of anchors and the set of sensors, respectively. The set of the neighbors of node $i$ in a given radius, $\mu_i$, is
\[\Phi(i,\mu_i)=\{r\in\Theta : d_{ri}<\mu_i\},\]
which may include anchors as well as sensors. In the classical DILOC algorithm, the following three deployment assumptions are required.
\begin{enumerate}
\item [\textbf{A1}] The locations of all the sensors lie inside the convex hull of the anchors, i.e., $Co\{\Omega\}\subseteq Co\{\Xi\}$.

\item [\textbf{A2}] For every sensor $i\in\Omega$, there are some $\mu_i>0$ such that a \emph{triangulation} set, $\mathscr{N}_i\subseteq\Phi(i,\mu_i)$, satisfying $|\mathscr{N}_i|=3$ exists, where $|\mathscr{N}_i|$ is the number of the elements in $\mathscr{N}_i$.

\item [\textbf{A3}] For any sensor $i\in\Omega$, there is a communication link between all nodes in the set $\{i\}\cup\mathscr{N}_i$.
\end{enumerate}

Finding the triangulation set $\mathscr{N}_{i}$ is an important step in the DILOC algorithm. Given the deployment density, sensor $i$ can set an initial communication radius $\mu_i$ to determine its neighbor set $\Phi(i,\mu_i)$. Then, sensor $i$ selects any three neighbors in $\Phi(i,\mu_i)$ and tests whether it is located in the convex hull of these neighbors. If all tests fail, sensor $i$ adaptively increases its communication radius in small increments. By repeating this process, the triangulation set will be finally determined.

Let $p_a=[p_1,p_2,p_3]^T$ and $p_s=[p_4,p_5,\ldots,p_n]^T$ be the locations of anchors and sensors, respectively. According to the representation of the barycentric coordinates of each node relative to its neighbors, we get
\begin{equation}\label{sys:2}
\begin{aligned}
  \begin{bmatrix}
    p_a \\
    p_s \\
  \end{bmatrix}=
          \begin{bmatrix}
            I_3 & \textbf{0} \\
            F & H \\
          \end{bmatrix}
  \begin{bmatrix}
    p_a \\
    p_s \\
  \end{bmatrix},
\end{aligned}
\end{equation}
where there are three nonzero elements in each row of $[F \ H]$. These are the barycentric coordinates of the nodes relative to their neighbors. In addition, the matrix $[F \ H]$ is row-stochastic and the block matrix $H$ is sub-stochastic. In particular, the row sums of matrix $H$ satisfy: if the neighbor set $\mathscr{N}_{i+3}$ of the sensor $i+3$ contains anchors, where $i=1,2,\ldots,n-3$, then $\sum_{j=1}^{n-3}H_{ij}<1$, and if there are no anchors in $\mathscr{N}_{i+3}$, then $\sum_{j=1}^{n-3}H_{ij}=1$.

The iteration of the DILOC algorithm in \cite{Khan2009Distributed} is written as
\begin{equation}\label{sys:3}
\tilde{p}_i(t+1)=\sum\nolimits_{r\in\mathscr{N}_{i}}a_{ir}\tilde{p}_r(t), \ i\in\Omega,
\end{equation}
where $\tilde{p}_i(t)$ is the location estimate of sensor $i$ at time $t$, and $a_{ir}$ are the barycentric coordinates of sensor $i$ relative to its three neighbors in $\mathscr{N}_{i}$. By the assumption \textbf{A3}, $a_{ir}$ are known to sensor $i$. The iteration process (\ref{sys:3}) can also be computed in another form \cite{Huang2018Asynchronous}, which can be written as
\begin{equation}\label{sys:4}
\tilde{p}_i(t+1)=\tilde{p}_i(t)+\gamma\sum\nolimits_{r\in\mathscr{N}_{i}}a_{ir}\big(\tilde{p}_r(t)-\tilde{p}_i(t)\big), \ i\in\Omega,
\end{equation}
where $\gamma$ is a constant. Obviously, (\ref{sys:4}) is a more general form, which covers (\ref{sys:3}) as a special case of $\gamma=1$. Define $\tilde{p}_s(t)=[\tilde{p}_4(t),\tilde{p}_5(t),\ldots,\tilde{p}_n(t)]^T$.
Eq. (\ref{sys:4}) can be expressed equivalently as the following compact form
\begin{equation}\label{sys:5}
\tilde{p}_s (t+1)=\gamma Fp_a+Q\tilde{p}_s(t),
\end{equation}
where $Q=(1-\gamma)I_{n-3}+\gamma H$. It has been shown in \cite{Khan2009Distributed,Khan2009Diland,Deghat2011Distributed,Zhu2011Distributed,Lin2017Distributed,Huang2018Asynchronous} that the final location estimations of the sensors updating through (\ref{sys:4}) or (\ref{sys:5}) are expressed as
\begin{equation}\label{sys:6}
\begin{aligned}
\lim_{t\rightarrow\infty}\tilde{p}_s (t)\!=\![I_{n-3}\!-\!H]^{-1}F p_a,
\end{aligned}
\end{equation}
which is also the exact locations of the sensors. $[I_{n-3}\!-\!H]^{-1}F$ is the matrix of the barycentric coordinates of the sensors in terms of the anchors.

\begin{remark}
For each sensor $i$, its three neighbors in $\mathscr{N}_{i}$ may contain both anchors and other sensors. Sensor $i$ can determine its barycentric coordinates relative to its three neighbors in $\mathscr{N}_{i}$ by measuring the Euclidean distance, thereby estimating its own location. Because the locations of anchors are known, sensor $i$ can localize itself in one single step if its three neighbors in $\mathscr{N}_{i}$ are all anchors. And if $\mathscr{N}_{i}$ contains location-unknown sensors, sensor $i$'s one-step location estimation is not its accurate location. The purpose of the DILOC algorithm is to determine the final location estimation of sensor $i$ completely by the locations of anchors through distributed iteration, so that accurate localization is achieved.
\end{remark}

\subsection{DoS attacks model}\label{section:3.2}

DoS attacks refer to a kind of attacks in which an attacker controls some or all components of the system. DoS attacks can affect the timeliness of information exchange, resulting in a loss of data \cite{Persis2015Input}. Without loss of generality, it is assumed that the attacker can attack the communication network in the changing active time period, which means the attacked communication channels can neither be used for sending nor receiving data. Next, the attacker needs to pause the attack activity and move to sleep time to save energy for the next attack.

\begin{figure}[htb]
  \centering
    \includegraphics[width=2in]{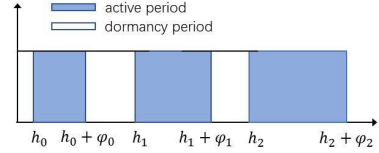}
    \vspace{-1ex}
  \caption{An example of attack schedules.}\label{fig2}
\end{figure}

Let $\{s_k\}_{k\in\mathbb{N}}$, where $s_0\geq0$, represents the time sequence of DoS attacks off/on conversion, i.e., the time instants when the attacker launches DoS attacks. Specifically, the attack instant $s_k$ and its dwell time $\varphi_k$ determine the $k$-th active period of DoS attacks, i.e., $S_k:=[s_k,s_k+\varphi_k]$, in which the attacker attacks parts or all of the communication channels, and then the information transmitted by the attacked channels will be lost. If $\varphi_k=0$, the $k$-th DoS takes the form of a single pulse at time $s_k$. Let
\begin{equation}\label{sys:7}
\mathbf{S}=\bigcup\nolimits_{k\in\mathbb{N}}S_k, \ \mathbf{M}=\mathbf{T}\backslash\mathbf{S}
\end{equation}
represent a set of time instants where communication is denied and allowed, respectively, where $\mathbf{T}=\{t : t=0,1,2\ldots\}$ is the set of all iteration time instants. To illustrate DoS attacks, a simple example is shown in Fig. \ref{fig2}.

\vspace{-1ex}

\vspace{-2ex}
\subsection{Simulations by the DILOC algorithm under DoS attacks }\label{section:3.3}

In the DILOC algorithm, each sensor always uses the received information from its neighbors to estimate the location. In other words, when a part of the communication links arriving at a certain sensor $i$ are attacked, the sensor $i$ estimates its location by using the information received from the remaining communication links. Define
\begin{equation}\label{sys:8}
\mathscr{N}^{D}_{i}(t)\!=\!\big\{j : j\!\in\!\mathscr{N}_{i} \ {\rm and} \ (j,i) \ {\rm is \ attacked \ at} \ t\!\in\!\mathbf{S} \big\}.
\end{equation}
For any $i\in\Omega$, the iteration form of the DILOC algorithm under DoS attacks is written as
\begin{small}
\begin{equation}\label{sys:9}
\tilde{p}_i(t+1)=\left\{
\begin{aligned}
&\tilde{p}_i(t)+\gamma\sum\nolimits_{r\in\mathscr{N}_{i}\backslash\mathscr{N}^{D}_{i}(t)}a_{ir}\big(\tilde{p}_r(t)-\tilde{p}_i(t)\big), t\in\mathbf{S},\\
&\tilde{p}_i(t)+\gamma\sum\nolimits_{r\in\mathscr{N}_{i}}a_{ir}\big(\tilde{p}_r(t)-\tilde{p}_i(t)\big), \ \ \ \ \ \ \ t\in\mathbf{M}.
\end{aligned}
\right.
\end{equation}
\end{small}
It is worth noting that the DILOC algorithm may not accurately locate the sensors due to the diversity of attack strategies of DoS, which are illustrated by the following example.

\begin{figure}[htb]
  \centering
    \includegraphics[width=1in]{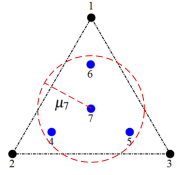}
    \vspace{-2ex}
  \caption{A network consisting of seven nodes, where $1,2,3$ and $4,5,6,7$ represent the anchors and the sensors, respectively.}\label{fig3}
\end{figure}

\begin{figure}[htb]
  \centering
    \subfigure[under attack strategy (\textbf{I})]{
    \includegraphics[width=1.7in]{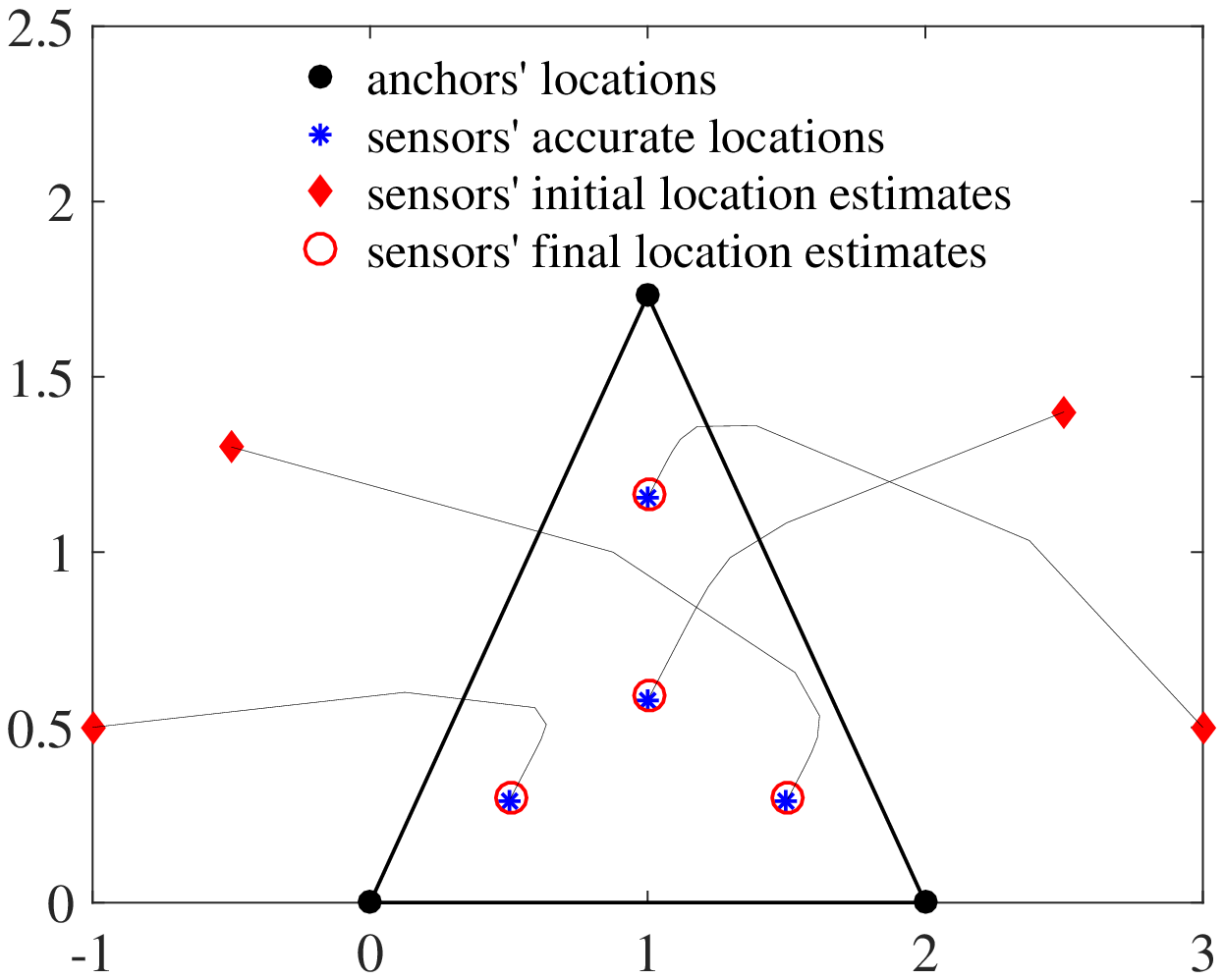}\label{fig4a}}\!\!\!\!\!\!\!\!\!
      \subfigure[attack strategy (\textbf{II})]{
    \includegraphics[width=1.7in]{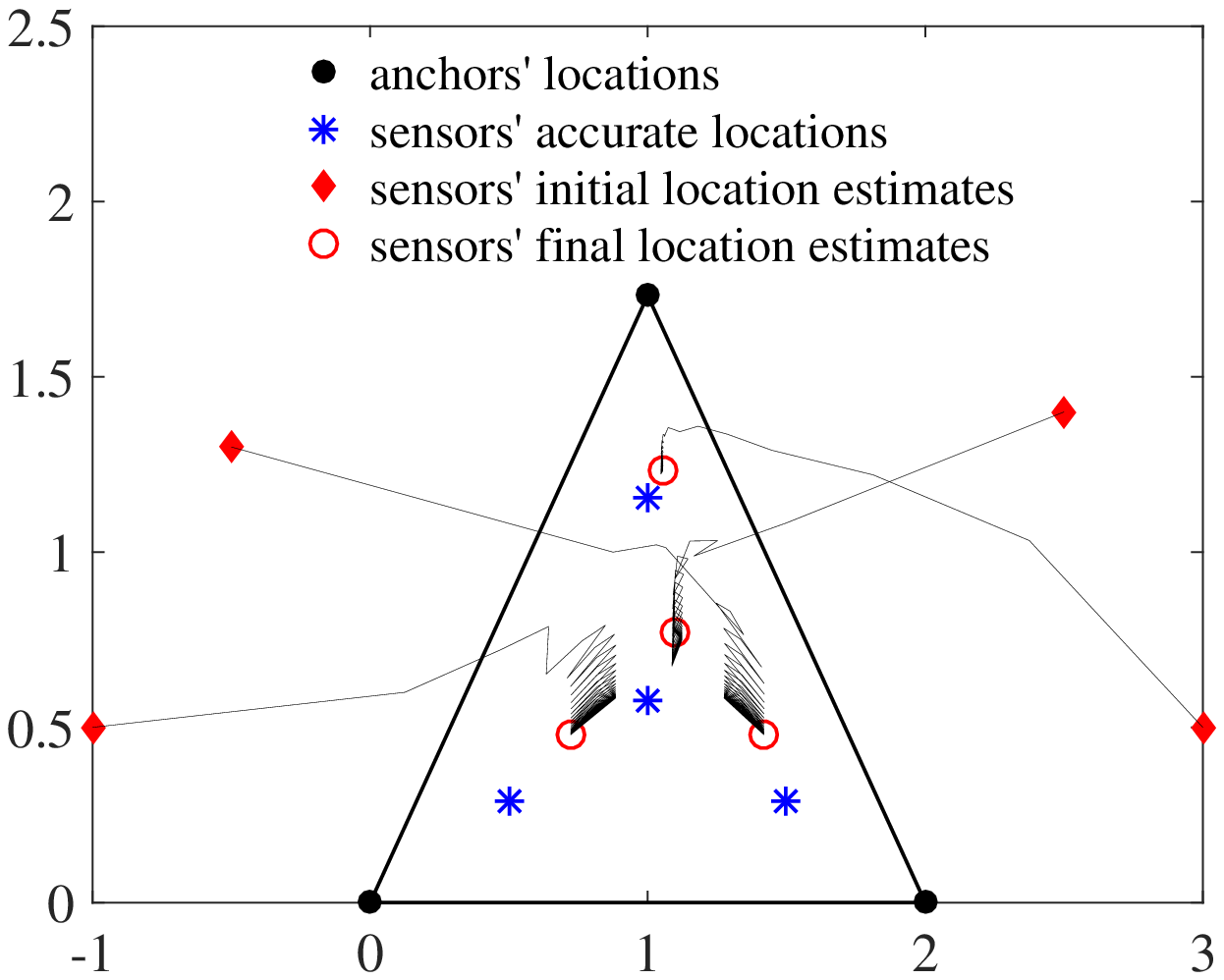}\label{fig4b}}
  \caption{Trajectories of the sensors' location estimations obtained by the DILOC algorithm under the DoS attack strategies (\textbf{I}) and (\textbf{II}). }\label{fig4}
\end{figure}

\begin{example}\label{example:1}
Consider a sensor network consisting of seven nodes, see Fig. \ref{fig3}, where the sets of anchors and sensors are $\Xi=\{1,2,3\}$ and $\Omega=\{4,5,6,7\}$, respectively. The locations of anchors are $p_1=(1,\sqrt{3}),p_2=(0,0),p_3=(2,0)$. The triangulation sets are chosen as $\mathscr{N}_4=\{2,5,7\}$, $\mathscr{N}_5=\{3,6,7\}$, $\mathscr{N}_6=\{1,4,7\}$, $\mathscr{N}_7=\{4,5,6\}$. We can see that there are no anchors in the sensor $7$'s triangulation set, and there is only one anchor in the triangulation sets of other sensors. Since no sensors communicate directly with all anchors, all sensors can not achieve accurate localization in one step. For each sensor $i\in\Omega$, its barycentric coordinates relative to its three neighbors in the triangulation set $\mathscr{N}_i$ can be calculated by using Euclidean distance among the nodes in the set $\{i\}\cup\mathscr{N}_i$ in the Cayley-Menger determinant. For example, we can use the distances $d_{74}=d_{75}=d_{76}=\frac{\sqrt{3}}{3}$, $d_{45}=d_{46}=d_{5,6}=1$ to compute $a_{74}=a_{75}=a_{76}=\frac{1}{3}$. Let the gain parameter be $\gamma=1/2$ in Eq. (\ref{sys:9}). Suppose the attacker attacks at time instants $s_k=3k$, $k\in\mathbb{N}$, and the dwell time is $\varphi_k=1$, $k\in\mathbb{N}$. We below consider two DoS attack strategies:
\begin{enumerate}
\item [(\textbf{I})] The attacker attacks communication links $(2,4)$, $(5,4)$, $(7,4)$ at time instants $3k$, $k\in\mathbb{N}$ and communication links $(4,7)$, $(5,7)$, $(6,7)$ at time instants $3k+1$, $k\in\mathbb{N}$.

\item [(\textbf{II})] The attacker attacks communication links $(2,4)$, $(3,5)$, $(6,7)$ at time instants $3k$, $k\in\mathbb{N}$ and communication links $(1,6)$, $(4,7)$, $(5,7)$ at time instants $3k+1$, $k\in\mathbb{N}$.
\end{enumerate}
It can be observed from Fig. \ref{fig4} that the DILOC algorithm can precisely determine the locations of the sensors under attack strategy (\textbf{I}), but not under attack strategy (\textbf{II}).
\end{example}

\vspace{-4ex}

\begin{remark}
As can be seen from Example \ref{example:1}, the DILOC algorithm cannot achieve accurate sensor localization under arbitrary attack strategy. In fact, the DILOC algorithm is only effective for one kind of attack strategy, that is, the attacker either attacks all three communication links to a node or does not attack any communication link to the node. If the attacker launches other attack strategies, the DILOC algorithm will fail. In addition, the effectiveness of DoS attacks is independent of the communication structure and initial location estimations.
\end{remark}

\subsection{AS-DILOC algorithm}\label{section:3.4}

In this paper, we propose a consensus-based distributed iterative algorithm which can accurately locate sensors under any DoS attack strategy under the deployment assumptions $\mathbf{A1}-\mathbf{A3}$, where each node estimates its location by using an  abandonment-strategy rule. The proposed algorithm is shown in Algorithm \ref{algorithm:1}.

\vspace{1ex}
\hrule
\vspace{0.5ex}
\noindent\textbf{Algorithm 1} \ AS-DILOC algorithm
\label{algorithm:1}
\vspace{0.5ex}
\hrule
\vspace{0.5ex}
\begin{enumerate}
\item [1:] Set the time sequence $\{s_k\}_{k\in\mathbb{N}}$ of DoS attacks off/on conversion, dwell time $\varphi_k$, $k\in\mathbb{N}$, and the initiate location estimates $\tilde{p}_i(0)$, $i=1,2,\ldots,n$;

\item [2:] For each location-unknown sensor $i\in\Omega$, the following update strategy is designed.
\begin{enumerate}
\item [I)] If sensor $i$ can receive the neighbors' information from all three communication links arriving at itself, then the following location estimate rule is designed:
\begin{equation}\label{sys:10}
\tilde{p}_i(t\!+\!1)\!=\!(1-\gamma)\tilde{p}_i(t)\!+\!\gamma\sum\nolimits_{r\in\mathscr{N}_{i}}a_{ir}\tilde{p}_r(t),
\end{equation}
\item [II)] If sensor $i$ detects that the information on at least one communication link arriving at it is intercepted, then a strategy of abandoning update is adopted, i.e.,
\begin{equation}\label{sys:11}
\tilde{p}_i(t\!+\!1)\!=\!\tilde{p}_i(t).
\end{equation}
\end{enumerate}

\item [3:] Advance the time to $t+1$, then go to step 2.
\end{enumerate}
\vspace{0.5ex}
\hrule
\vspace{1.5ex}

\begin{remark}
In the DILOC algorithm (\ref{sys:9}), if some communication links arriving at node $i$ are attacked by the attacker, namely, node $i$ cannot receive the neighbor information on these communication links, then node $i$ will use the neighbor information received from the remaining communication links to estimate its location. The difference between the AS-DILOC algorithm and the classic DILOC algorithm is reflected in the strategy of processing neighbor information when the sensor network is attacked. For example, in the AS-DILOC algorithm, even if only one communication link arriving at node $i$ is attacked, node $i$ adopts a strategy of abandoning update, i.e., $\tilde{p}_i(t+1)=\tilde{p}_i(t)$.
\end{remark}

\vspace{-1ex}
\section{Convergence analysis}\label{section:4}

The global convergence of the AS-DILOC algorithm is analyzed by using the sub-stochastic matrix and the composition of binary relations.

We first convert Eqs. (\ref{sys:10}) and (\ref{sys:11}) to a matrix-vector form:
\begin{equation}\label{sys:12}
\begin{aligned}
\tilde{p}_s (t\!+\!1)\!=\!\gamma F(t)p_a\!+\!\big[I_{n-3}\!-\!\gamma \big(N(t)\!-\!H(t)g\big)\big]\tilde{p}_s(t),
\end{aligned}
\end{equation}
where \begin{small}$N(t)={\rm diag}\{\sum_{r}[F(t) \ H(t)]_{1r},\ldots,\sum_{r}[F(t) \ H(t)]_{n-3,r}\}$,\end{small}
\begin{equation*}
\begin{aligned}
&[F(t) \ H(t)](i,:)\!=\!\textbf{0}, \ \ \ \ \ \ \ \ \ \ {\rm if} \ t\in\mathbf{S}, \mathscr{N}^{D}_{i}(t)\!\neq\!\emptyset,\\
&[F(t) \ H(t)](i,:)\!=\![F \ H](i,:), {\rm if} \ t\in\mathbf{M} \ {\rm or} \ t\!\in\!\mathbf{S},\mathscr{N}^{D}_{i}\!(t)\!\neq\!\emptyset,
\end{aligned}
\end{equation*}
for $i=1,2,\ldots,n-3$. Let $Q(t)=I_{n-3}\!-\!\gamma \big(N(t)-H(t)\big)$, then the solution of iterative process (\ref{sys:12}) can be written as
\begin{equation}\label{sys:13}
\begin{aligned}
\lim_{t\rightarrow\infty}\tilde{p}_s (t)=\prod\nolimits_{t=0}^{\infty}Q(t)\tilde{p}_s(0)+\lim_{t\rightarrow\infty}\Gamma(t)p_a,
\end{aligned}
\end{equation}
where $\Gamma(t)=\gamma F(t)+\gamma\sum_{j=1}^{t}\prod_{i=j}^{t}Q(i)F(j-1)$.

We observe from (\ref{sys:5}) that the final location estimations of the sensors are located in the convex hull of the anchors and linearly represented by the locations of the anchors. Therefore, the following equation should be proved first
\begin{equation}\label{sys:14}
\begin{aligned}
\prod\nolimits_{t=0}^{\infty}Q(t)=\textbf{0}.
\end{aligned}
\end{equation}
Furthermore, in order to ensure that the final location estimations of sensors are the same with the accurate locations, it is also necessary to prove that
\begin{equation}\label{sys:15}
\begin{aligned}
\lim_{t\rightarrow\infty}\Gamma(t)=\![I_{n-3}\!-\!H]^{-1}F.
\end{aligned}
\end{equation}
Our next major task is to give detailed proofs of Eqs. (\ref{sys:14}) and (\ref{sys:15}), respectively. Before proceeding, we construct a digraph  $\mathscr{G}[M(t)]=\big\{\mathscr{V}[M(t)],\mathscr{E}[M(t)]\big\}$, $t\in\mathbb{N}$, where
\begin{equation*}
\begin{aligned}
M(t)=\begin{bmatrix}
            1 & \textbf{0} \\
            \gamma F(t)\textbf{1} & Q(t) \\
          \end{bmatrix}
\end{aligned}
\end{equation*}
is the corresponding weighted adjacency matrix, $\mathscr{V}[M(t)]=\{0,1,\ldots,n-3\}$ is the node set in which the elements are the row indexes of matrix $M(t)$ except the first one which is labeled as 0, and $\mathscr{E}[M(t)]$ is the set of arcs.

We first present the result of directed edges composition to prove Eqs. (\ref{sys:14}) and (\ref{sys:15}).

\begin{theorem}\label{theorem:1}
Consider the AS-DILOC algorithm. Under the deployment assumptions $\mathbf{A1}-\mathbf{A3}$, if the parameter $\gamma$ is selected within the interval $(0,1)$, then for any time intervals $[s_k,s_{k+P})$, $k\in\mathbb{N}$, where $P\!=\!\max\{d_{i\rightarrow j}: i\!\in\!\Xi, j\!\in\!\Omega\}$, we have
\begin{eqnarray}\label{sys:17}
\begin{aligned}
(0,i)\!\in\!\mathscr{E}\big[M(s_k)\big]\!\circ\!\cdots\!\circ\!\mathscr{E}\big[M(s_{k+P}\!-\!1)\big],
\end{aligned}
\end{eqnarray}
for $i=1,2,\ldots,n-3$.
\end{theorem}

\begin{IEEEproof}
The communication links among the nodes are described by a digraph $\mathscr{G}$. It is known from \cite{Huang2018Asynchronous} that there is a directed path from the anchor set to each location-unknown sensor in the digraph $\mathscr{G}$ under the assumptions $\mathbf{A1}-\mathbf{A3}$. Assume that the directed path from the anchors set $\Xi$ to each sensor $i_z+3$ is described as $\mathscr{P}_{a\rightarrow i_z+3}\!=\!(a,i_1\!+\!3),(i_1\!+\!3,i_2\!+\!3),\ldots,(i_{z-1}\!+\!3,i_z\!+\!3)$,
where $a\in\Xi$ and $i_1+3,i_2+3,\ldots,i_{z}+3\in\Omega$ with $i_1,i_2,\ldots,i_z\in\{1,2,\ldots,n-3\}$.

We first consider the arc $(a,i_1+3)$ and the time interval $[s_k,s_{k+1})$. Since the communication channels reaching node $i_1+3$ may be attacked in the interval $[s_k,s_{k}+\varphi_k)$, so the arc $(a,i_1+3)$ may not belong to the digraph $\mathscr{G}$ at time $t\in[s_k,s_{k}+\varphi_k)$. And since $[s_{k}+\varphi_k,s_{k+1})$ is the dormant period of nodes, node $i_1+3$ can get information about all its neighbors at any $t\in[s_{k}+\varphi_k,s_{k+1})$. According to the definition of $[F(t) \ H(t)]$, we can get $\sum_{j}[H(s_{k}\!+\!\varphi_k)]_{i_1j}<1$, equivalently, $\sum_{j}[Q(m\delta\!+\!T_d)]_{i_1j}<1$. This implies that $(0,i_1)\in\mathscr{E}[M(m\delta\!+\!T_d)]$. Furthermore, it can be observed that the row-stochastic matrix $M(t)$ has diagonal elements under the condition $0<\gamma<1$, namely, the digraph $\mathscr{G}[M(t)]$ has self-loops on all nodes $0,1,\ldots,n-3$. Consequently,
\begin{equation}\label{sys:17}
\begin{aligned}
(0,i_1)\in\mathscr{E}\big[M(s_{k})\big]\circ\cdots\circ\mathscr{E}\big[M(s_{k+1}\!-\!1)\big],
\end{aligned}
\end{equation}
where $(0,0)\!\in\!\mathscr{E}[M(s_{k})],\ldots,(0,0)\!\in\!\mathscr{E}[M(s_{k}\!+\!\varphi_k\!-\!1)]$, $(0,i_1)\!\in\!\mathscr{E}[M(s_{k}\!+\!\varphi_k)],
(i_1,i_1)\!\in\!\mathscr{E}[M(s_{k}\!+\!\varphi_k\!+\!1)],\ldots,(i_1,i_1)\!\in\!\mathscr{E}[M(s_{k+1}\!-\!1)]$.

Consider the arc $(i_{r}\!+\!3,i_{r+1}\!+\!3)$ and the attack period $[s_{k+r},s_{k+r+1})$, where $r=1,2,\ldots,z-1$. Obviously, each node can get information about all its neighbors during the dormant period $[s_{k+r}\!+\!\varphi_{k+r},s_{k+r+1})$. From the definition of $[F(t) \ H(t)]$, we have $[H(s_{k+r}\!+\!\varphi_{k+r})]_{i_{r+1}i_r}>0$, and further $[Q(s_{k+r}\!+\!\varphi_{k+r})]_{i_{r+1}i_r}>0$. This means that $(i_r,i_{r+1})\in\mathscr{E}[M(s_{k+r}\!+\!\varphi_{k+r})]$. It thus follows that
\begin{equation}\label{sys:18}
\begin{aligned}
(i_{r},i_{r+1})\!\in\!\mathscr{E}\big[M(s_{k+r})\big]\!\circ\!\cdots\!\circ\!\mathscr{E}\big[M(s_{k+r+1}\!-\!\!1)\big],
\end{aligned}
\end{equation}
where $(i_{r},i_{r})\!\in\!\mathscr{E}[M(s_{k+r})],\ldots,(i_{r},i_{r})\!\in\!\mathscr{E}[M(s_{k+r}\!+\!\varphi_{k+r}\!-\!1)],
(i_{r},i_{r+1})\!\in\!\mathscr{E}[M(s_{k+r}\!+\!\varphi_{k+r})],(i_{r+1},i_{r+1})\!\in\!\mathscr{E}[M(s_{k+r}\!+\!\varphi_{k+r}\!+\!1)],\ldots,
(i_{r+1},i_{r+1})\!\in\!\mathscr{E}[M(s_{k+r+1}\!-\!1)]$.

Combining (\ref{sys:17}) and (\ref{sys:18}), it can be obtained that
\begin{equation}\label{sys:19}
\begin{aligned}
(0,i_z)\!\in\!\mathscr{E}\big[M(s_k)\big]\!\circ\!\cdots\!\circ\!\mathscr{E}\big[M(s_{k+z}\!-\!1)\big],
\end{aligned}
\end{equation}
where $z$ is the length of the path $\mathscr{P}_{a\rightarrow i_z+3}$ and it satisfies $z\leq P$. In addition, the fact that the nodes have self-loops in the digraph $\mathscr{G}[M(t)]$, $t\in\mathbb{N}$ guarantees that
\begin{equation}\label{sys:20}
\begin{aligned}
(i_z,i_z)\!\in\!\mathscr{E}\big[M(s_{k+z})\big]\!\circ\!\cdots\!\circ\!\mathscr{E}\big[M(s_{k+P}\!-\!1)\big].
\end{aligned}
\end{equation}
By (\ref{sys:17}) and (\ref{sys:20}), we have
\begin{equation}\label{sys:21}
\begin{aligned}
(0,i_z)\!\in\!\mathscr{E}\big[M(s_k)\big]\!\circ\!\cdots\!\circ\!\mathscr{E}\big[M(s_{k+P}\!-\!1)\big].
\end{aligned}
\end{equation}
Equivalently, we have $(0,i)\in\mathscr{E}[M(s_k)]\circ\cdots\circ\mathscr{E}[M(s_{k+P}\!-\!1)]$ for any $i=1,2,\ldots,n-3$.
\end{IEEEproof}

Based on the conclusion of Theorem \ref{theorem:1}, we analyze Eq. (\ref{sys:14}) and Eq. (\ref{sys:15}) in detail in Theorem \ref{theorem:2}.

\begin{theorem}\label{theorem:2}
Under the deployment assumptions $\mathbf{A1}-\mathbf{A3}$, if the parameter $\gamma$ is selected within the interval $(0,1)$, then Eq. (\ref{sys:14}) and Eq. (\ref{sys:15}) can be derived, that is, the AS-DILOC algorithm can converge to the exact sensor locations under DoS attacks.
\end{theorem}

\begin{IEEEproof}
For any time interval $[s_{mP},s_{(m+1)P})$, where $m\in\mathbb{N}$, we know from Theorem \ref{theorem:1} that $(0,i)\in\mathscr{E}[M(s_{mP})]\circ\cdots\circ\mathscr{E}[M(s_{(m+1)P}-1)]$, $i=1,2,\ldots,n-3$. Let $\delta_m$ denote the number of time instants in the interval $[s_{mP},s_{(m+1)P})$. According to Definition \ref{difinition:2}, there are a series of nodes $i_1,i_2,\ldots,i_{\delta}\in\{1,2,\ldots,n-3\}$ such that $(0,i_1)\in\mathscr{E}[M(s_{mP})],(i_1,i_2)\in\mathscr{E}[M(s_{mP}+1)],\ldots, (i_{\delta_m-1},i_{\delta_m})\in\mathscr{E}[M(s_{(m+1)P}-1)]$, where $i_{\delta_m}=i$ and there may be duplicate elements in the nodes set $\{i_1,i_2,\ldots,i_{\delta_m}\}$.

Below we analyze the matrices product $\prod_{t=s_{mP}}^{s_{(m+1)P}-1}Q(t)$ associated with the interval $[s_{mP},s_{(m+1)P})$. Define
\begin{eqnarray}\label{sys:22}
\begin{aligned}
\sigma=\min\big\{\gamma a_{ir}, 1-\gamma\sum\nolimits_{r\in\mathscr{N}_i}a_{ir} : i,r=1,2,\ldots,n\big\},
\end{aligned}
\end{eqnarray}
then all elements of $M(t)$ are greater than or equal to $\sigma$. By $(0,i_1)\in\mathscr{E}[M(s_{mP})]$, we have $[\gamma F\mathbf{1}]_{i_1}\geq\sigma$, it follows that
\begin{equation}\label{sys:23}
\begin{aligned}
\sum\nolimits_{j=1}^{n-3}[Q(s_{mP})]_{i_1j}=1-[\gamma F\mathbf{1}]_{i_1}\leq1-\sigma<1.
\end{aligned}
\end{equation}
For any $q=1,2,\ldots,\delta_m-1$, since $(i_q,i_{q+1})\in\mathscr{E}[M(s_{mP}+q)]$, we get $[Q(s_{mP}+q)]_{i_{q+1}i_q}\geq\sigma$. It thus follows that
\begin{small}
\begin{align}\label{sys:24}
&\Big[\prod\nolimits_{t=s_{mP}+1}^{s_{(m+1)P}-1}Q(t)\Big]_{i_{\delta_m} i_1}\nonumber\\
&\geq\!\big[Q(s_{(m+1)P}\!-\!1)\big]_{i_{\delta_m} i_{\delta_m-1}}\!\!\cdots\!\!\big[Q(s_{mP}\!+\!1)\big]_{i_2 i_{1}}\!\geq\!\sigma^{\delta_m-1}.
\end{align}
\end{small}
Combining with (\ref{sys:23}) and (\ref{sys:24}), we can further deduce that
\begin{small}
\begin{align}\label{sys:25}
&\sum_{j=1}^{n-3}\Big[\prod\nolimits_{t=s_{mP}}^{s_{(m+1)P}-1}\!Q(t)\Big]_{i_{\delta_m}j}\nonumber\\
&=\!\!\sum_{j_1=1\atop j_1\neq i_1}^{n-3}\!\!\!\Big[\prod\nolimits_{t=s_{mP}+1}^{s_{(m+1)P}-1}\!\!\!Q(t)\Big]_{i_{\delta_m} j_1}\sum_{j=1}^{n-3}[Q(s_{mP})]_{j_1j}\nonumber\\
& \ \ \ \ +\Big[\prod\nolimits_{t=s_{mP}+1}^{s_{(m+1)P}-1}\!\!\!Q(t)\Big]_{i_{\delta_m} i_1}\sum_{j=1}^{n-3}[Q(s_{mP})]_{i_1j}\nonumber\\
&\!\!\leq(1-\sigma^{\delta_m-1})+\sigma^{\delta_m-1}(1-\sigma)=1-\sigma^{\delta_m}<1.
\end{align}
\end{small}

\vspace{-4ex}
\noindent
Equivalently, we have
\begin{equation}\label{sys:26}
\begin{aligned}
\Big\|\prod\nolimits_{t=s_{mP}}^{s_{(m+1)P}-1}\!\!\!Q(t)\Big\|_{\infty}\leq1-\sigma^{\delta_m}\leq1-\sigma^{\hat{\delta}}<1
\end{aligned}
\end{equation}
for any $m\in\mathbb{N}$, where $\hat{\delta}=\max\{\delta_m : m\in\mathbb{N}\}$. Consequently,
\begin{small}
\begin{align}\label{sys:27}
\Big\|\prod\nolimits_{t=0}^{\infty}\!Q(i)\Big\|_{\infty}
\!&\leq\!\prod\nolimits_{m=0}^{\infty}\Big\|\prod\nolimits_{t=s_{mP}}^{s_{(m+1)P}-1}\!\!Q(t)
\Big\|_{\infty}\Big\|\prod\nolimits_{t=0}^{s_0-1}\!\!Q(t)\Big\|_{\infty}\nonumber\\
\!&\leq\lim_{m\rightarrow\infty}\big(1-\sigma^{\hat{\delta}}\big)^m=0.
\end{align}
\end{small}
This guarantees the establishment of Eq. (\ref{sys:14}).

In the DILOC algorithm, it has been proved that the matrix $H$ is invertible if there is a directed path from the anchors set to each location-unknown sensor, thus $(I_{n-3}\!-\!H)^{-1}$ exists. According to the definitions $F(t)$ and $Q(t)$, we deduce that
\begin{equation}\label{sys:28}
\begin{aligned}
\big(I_{n-3}\!-\!Q(t)\big)(I_{n-3}\!-\!H)^{-1}F\!=\!\gamma F(t).
\end{aligned}
\end{equation}
By substituting (\ref{sys:28}) into $\Gamma(t)$, we obtain that
\begin{small}
\begin{align}
\lim_{t\rightarrow\infty}\!\!\Gamma(t)\!&=\!\!\lim_{t\rightarrow\infty}\!\!\Big(\!\sum_{i=0}^{t}Q(t)\!\cdots \! Q(i\!+\!1)\!\big(I_{n-3}\!-\!Q(i)\big)\!\Big)(I_{n-3}\!-\!\!H)^{-1}\!F\nonumber\\
&=\!\lim_{t\rightarrow\infty}\Big(I_{n-3}-Q(t)Q(t-1)\cdots Q(0)\Big)(I_{n-3}\!-\!H)^{-1}F\nonumber\\
&=\!(I_{n-3}\!-\!H)^{-1}F.
\end{align}
\end{small}
This also means that (\ref{sys:15}) is established.
\end{IEEEproof}

\begin{figure}[htb]
  \centering
    \subfigure[under attack strategy (\textbf{I})]{
    \includegraphics[width=1.7in]{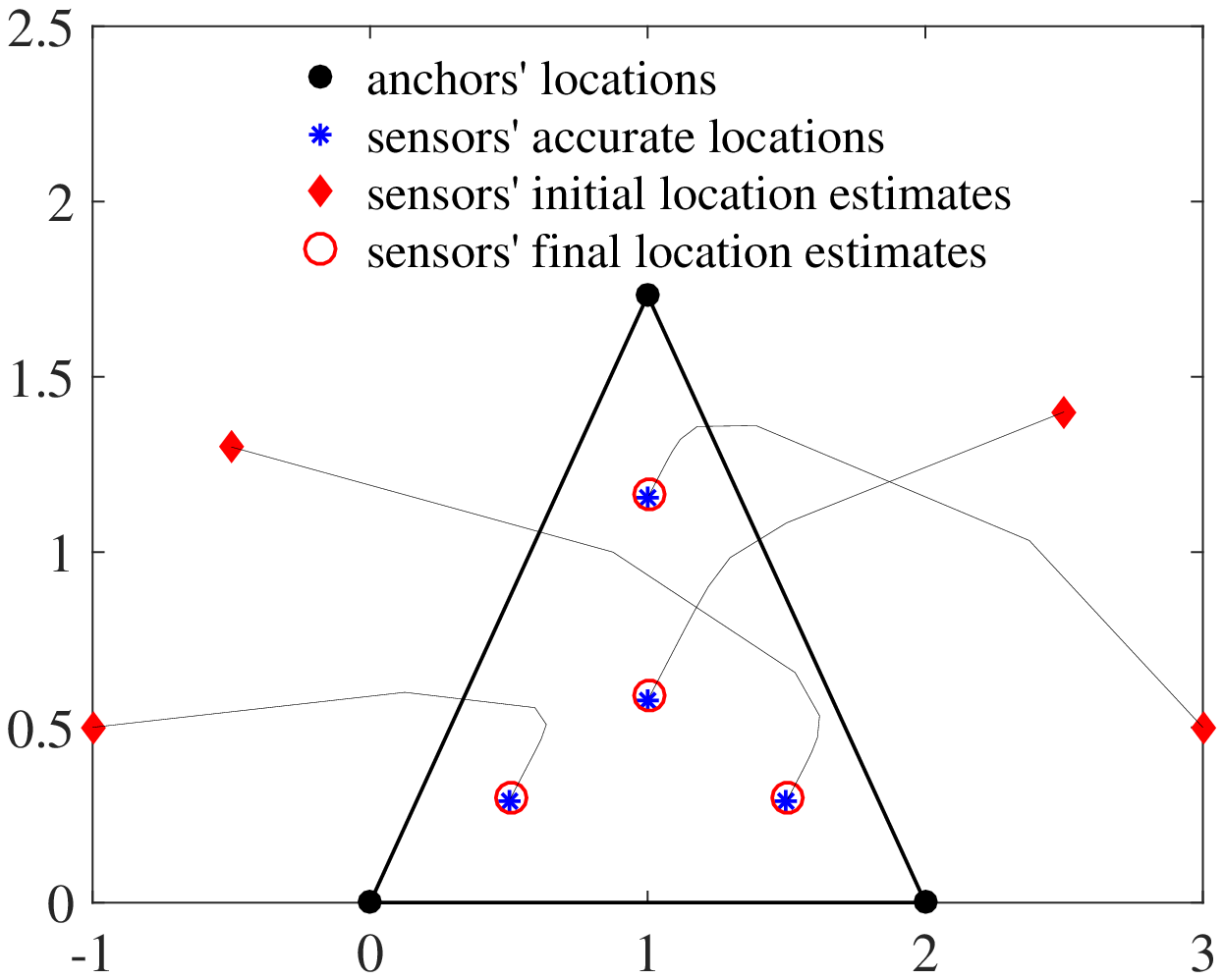}\label{fig5a}}\!\!\!\!\!\!\!\!\!
      \subfigure[under attack strategy (\textbf{II})]{
    \includegraphics[width=1.7in]{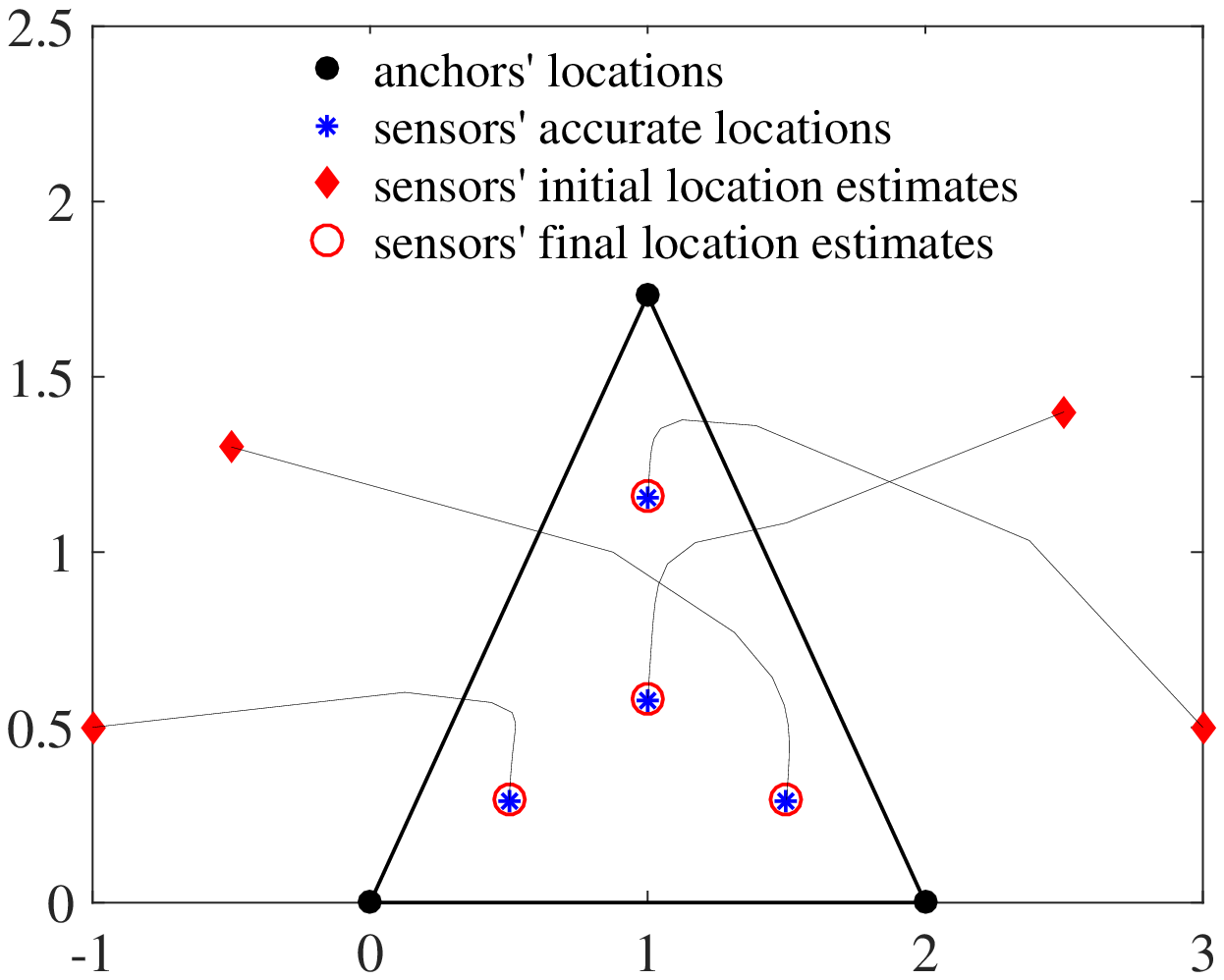}\label{fig5b}}
  \caption{Trajectories of the sensors' location estimates obtained by the AS-DILOC algorithm under the DoS attack strategies (\textbf{I}) and (\textbf{II}) in Example \ref{example:1}.}\label{fig5}
\end{figure}

\begin{example}\label{example:2}
Consider a sensor network consisting of the anchor set $\Xi=\{1,2,3\}$ and the sensor set $\Omega=\{4,5,6,7\}$ in the 2-dimensional Euclidean space. The triangulation sets and location coordinates of the sensors are set the same as Example \ref{example:1}. Let the gain parameter be $\gamma=1/2$ in Eq. (\ref{sys:10}). It can be observed from Fig. \ref{fig5} that the AS-DILOC algorithm can precisely determine the locations of the sensors whether under attack strategies (\textbf{I}) or (\textbf{II}) considered in Example \ref{example:1}.
\end{example}

\begin{remark}\label{remark:3}
The main contribution of this paper is to propose an AS-DILOC algorithm based on barycentric coordinates measurement in the situation of DoS attacks. We show through Example \ref{example:1} that the DILOC algorithm may fail when the attacker changes the attack strategy. We prove that the AS-DILOC algorithm proposed in this paper can achieve accurate sensor localizations under arbitrary attack strategy, which is illustrated by Example \ref{example:2}.
\end{remark}

\section{Conclusion}\label{section:6}

The problem of determining the sensors locations in the presence of DoS attacks has been studied in this paper. The AS-DILOC algorithm has been proposed based on the barycentric coordinates which only involve relative distance measurement. The global convergence of the AS-DILOC algorithm has been analyzed by using the sub-stochastic matrix and the composition of binary relations. Moreover, the effectiveness of the AS-DILOC algorithm under DoS attacks has been verified by a simulation example.

\section*{acknowledgment}

The authors gratefully acknowledge the suggestions and comments by the associate editor and anonymous.

\end{document}